\documentclass{aa} 

\usepackage{graphicx}
\usepackage{txfonts}

\begin{document}

\title{Diagnostics of inhomogeneous stellar jets}
      \subtitle{Convolution effects and data reconstruction}
\author{
F.~De~Colle \inst{1},
C.~del~Burgo \inst{1}
\& A.~C.~Raga  \inst{2}}

\offprints{F. De Colle}

\titlerunning{Diagnostics of inhomogeneous stellar jets}
\authorrunning{F. De Colle et al.}

\institute{
Dublin Institute for Advanced Studies (DIAS), 31 Fitzwilliam Place, Dublin 2, Ireland \\
\email{fdc, cburgo@cp.dias.ie}\\
\and
Instituto de Ciencias Nucleares, Universidad Nacional Aut\'onoma de M\'exico, Ap.P. 70543, 04510 DF, Mexico\\
\email{raga@nucleares.unam.mx}
}


\abstract
   {
    In the interpretation of stellar jet observations, the 
    physical parameters are usually determined
    from emission line ratios, obtained from spectroscopic observations
    or using the information contained in narrow band images.
    The basic hypothesis in the interpretation of the
    observations is that the emitting region is homogeneous
    along the line of sight.
    Actually, stellar jets are in general not homogeneous,
    and therefore line of sight convolution effects may lead to 
    the main uncertainty in the determination of the physical
    parameters.}
   {
    This paper is aimed at showing the systematic errors 
    introduced when assuming an homogeneous medium, and studying
    the effect of an inhomogeneous medium on plasma
    diagnostics for the case of a stellar jet.
    In addition, we explore how
    to reconstruct the volumetric physical parameters
    of the jet (i.~e., with dependence both across and along
    the line of sight).}
   {We use standard techniques to determine 
    the physical parameters, i.~e., the electron density,
    temperature and hydrogen ionisation fraction across the jet,
    and a multi-Gaussian method to invert the Abel transform
    and determine the reconstructed physical structure.
      }
   {When assuming an homogeneous medium the physical parameters, 
    integrated along the line of sight,
    do not represent the average of the true values, and do not have
    a clear physical interpretation. We show that when
    some information is available on the emissivity profile
    across the jet, it is then possible to obtain appropriate derivations
    of the electron density, temperature and ionisation fraction.}
   {}

\keywords{ISM: Herbig-Haro objects -- 
          ISM: jets and outflows --
          Techniques: image processing --
          Methods: data analysis --
          Stars: pre-main sequence --
          winds, outflows
         }

\maketitle


\section{Introduction}

Stellar outflows are generally observed as chains of bright knots
with a characteristic spectrum that mainly includes
forbidden emission lines
(see, e.~g., the review by Reipurth \& Bally \cite{rb01}).

Different approaches are used to extract the information on the 
outflow excitation conditions from the observed emission lines.
To calculate the electron density $n_\mathrm{e}$ the 
standard method involves the line ratio 
$[\rm{SII}] \lambda6716/ \lambda6731$, which is
strongly dependent on $n_\mathrm{e}$ and less sensitive
to the electron temperature $T$ (see, e.~g. Osterbrock \cite{o89}).
The [SII] ratio, together with the ratios between 
lines with different excitation temperatures
(e.~g. $[\rm{OI}] \lambda5577/ \lambda6300$)
is used to determine both $n_\mathrm{e}$ and $T$.
Moreover, the [SII],  
$[\rm{OI}] \lambda\lambda 6300, 6363$ and
$[\rm{NII}] \lambda\lambda 6548, 6583$ emission lines
are often used to calculate, apart from $n_\mathrm{e}$
and $T$, the hydrogen ionisation fraction 
x$_\mathrm{H}$ and the total density 
n$_\mathrm{H}$ (=$n_\mathrm{e}/x_\mathrm{H}$)
(Bacciotti \& Eisl\"offel \cite{be99}, hereafter BE99).
Additionally, several line ratios in the optical and near infrared 
wavelength range
(e.~g. Brugel et al. \cite{b81}; Pesenti et al. \cite{p03};
Nisini et al. \cite{n05}; Podio et al. \cite{p06}) can be combined,
with weights representing their respective errors
(Hartigan \& Morse \cite{hm07}, hereafter HM07), to determine the
physical parameters of the HH objects.

Another possibility
consists of obtaining predictions of emission line intensities
from plane parallel shock models (e.~g. Hartigan et al. \cite{h87}, 
Hartigan et al. \cite{h94}, Lavalley-Fouquet et al. \cite{l00}, 
Pesenti et al. \cite{p03}) or
from numerical simulations that include a
detailed treatment of the ionisation and recombination
evolution of several different species (e.~g. Raga \cite{r07}),
and compare the results with observations.

It is important to develop strategies to properly extract
the maximum possible amount of information from the observed
line ratios.
Convolution, instrumental and
observing conditions (i.~e. seeing) affect
the derivation of physical parameters from observations.
Projection or convolution effects are an obvious consequence
of the lack of information on the
spatial distribution along the line of sight.
Additionally, if the cooling region is not resolved the 
observed emission lines come from regions
with different excitation conditions
\emph{along} the main jet axis.
Actually, also in observations where the cooling region 
is well resolved the determined parameters are a convolution 
of the volumetric parameters along the line of sight, i.~e.
\emph{across} the jet.

Only for an homogeneous medium
the mean values of $n_\mathrm{e}$, $T$, $x_\mathrm{H}$
(integrated within the beam size and along the line of sight) will
coincide with the corresponding values calculated from the
observed line ratios.
In the case of an inhomogeneous medium it is necessary
to understand what relation holds between the derived
and the volumetric mean physical parameters.

This problem has been studied in detail 
in the context of stellar atmospheres by several authors 
(e.~g. Doschek \cite{d84}, Almleaky et al. \cite{a89}, 
hereafter ABS89, Brown et al. \cite{b91}, Judge et al. \cite{j97},
McIntosh et al. \cite{m98}, Judge \cite{j00}).
As they show the information contained
in a set of line ratios may be used to determine
the function $\zeta(n_\mathrm{e})$, defined as the ``emission measure
differential in density'', given by 
$n_\mathrm{e}^2(z) dz=\zeta(n_\mathrm{e}) dn_\mathrm{e}$
(e.~g. Judge \cite{j00}),
where $z$ is the position across the stellar atmosphere.
If several line ratios are available, and 
if the geometry of the system is known, it becomes possible to infer
the dependence of the density $n_\mathrm{e}$ along the line of sight.

Despite its importance, this problem has not been studied in similar
detail in the context of stellar outflows.
As far as we are aware, only Safier (\cite{s92}) studied the influence of 
inhomogeneities on the density diagnostics in outflows from T Tauri stars,
applying a simple model of an isothermal medium
with a power law radial dependence for the electron and total densities,
and showing that different ratios may be 
used to obtain information on the density structure.

Recently, observations with information
on emission profiles across the jet 
have been presented by several authors
(e.~g. Beck et al. \cite{bec07}, Bacciotti et al. \cite{b00}, HM07, Coffey et al. \cite{c07}).
If the emission line profiles are limited to
a few observational values across the jet
we will show that this information can be used 
to derive the values of 
the physical parameters $n_\mathrm{e}$, $T$, and $x_\mathrm{H}$
assuming a Gaussian profile for the emission lines.
If the emission profiles across the jet
are well sampled, we will show in this paper how standard 
tomographic techniques (e.~g. Craig \& Brown \cite{cb86}, Brown \cite{b95})
may be applied to reconstruct the volume emission line intensities
(i. e., the emission coefficients) 
from the values integrated along lines of sight, which can
then be used to determine the $n_\mathrm{e}$, $T$ and $x_\mathrm{H}$ cross
sections of the jet.

The paper is organised as follows. In Section 2, the 
effect of stratification on the interpretation of the electron
density, temperature and
hydrogen ionisation diagnostics is discussed.
Section 3 presents a simple technique
for obtaining information on the radial structure of the 
jet, and an example of its application to data of the HH30 jet.
Finally, Section 4 summarises the obtained results.


\section{Convolution effects on jets}

\subsection{Convolution effects for a non-resolved jet cross section}

\begin{figure}
\centering
\includegraphics[width=6cm]{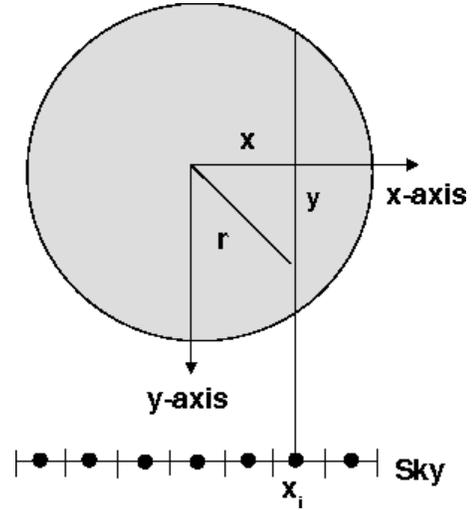}
\caption{Schematic representation of a cross section of a stellar jet.
         The jet is moving in the direction perpendicular to
         the plane of the image (the $z$-axis, not shown in the 
         Figure), and is assumed to be axisymmetric.
         The observer is
         looking at the jet along the $y$-direction, and
         the observations are represented by a series of points
         along the $x$-axis, each one corresponding to the pixel
         size and obtained integrating 
         the volume intensities along the line of sight
         and on the beaming size $\Delta x \Delta z$.}
\label{fig1}
\end{figure}

We consider an axisymmetric jet in a
three - dimensional cartesian reference frame.
The $z$-axis of this system is taken along with the main axis of the jet.
The observer is looking at the jet along the $y$-axis, and
the $xz$-plane coincides with the plane of the sky (see Fig. \ref{fig1}).
The cylindrical radius $r$ is the distance between an arbitrary
point and the $z$-axis.

The jet is observed in the $[\rm{SII}] \lambda\lambda
6716, 6731$, $[\rm{OI}] \lambda\lambda 6300, 6363$ and
$[\rm{NII}] \lambda\lambda 6548, 6583$ emission lines.
These lines give three ratios which depend on the three variables
to be determined: $n_\mathrm{e}$, $T$, $x_H$.
These emission line intensities
are observed at a certain number of points $x_i$ 
(corresponding to the pixel size)
in the plane of the sky.

In the following, capital letters will be used to design
quantities integrated and convolved (according to the nature of the 
variable) along the line of sight, and
lower case letters for volume quantities.
For instance, $i$ represents the volume emissivity
(defined by eq. \ref{eq:int}),
while $I$ is the intensity (i.~e. the emissivity 
\emph{integrated} along the line of sight,
with units: erg s$^{-1}$ cm$^{-2}$ sr$^{-1}$);
$n_\mathrm{e}$ represents the electron density across the jet,
while $N_\mathrm{e}$ is the electron density obtained \emph{convolving}
$n_\mathrm{e}$ on a volume element.
Only the electron temperature is designed by $T$ in both cases.

To clarify the effect of inhomogeneities
on plasma diagnostics, first some results
from ABS89 are reviewed.
The ratio $G_{12}$ between two observed
line luminosities is given by:
\begin{equation}
  G_{12} = \frac{\int i_1 dV}{\int i_2 dV} ,
  \label{eq:g12}
\end {equation}
where $dV$ is the volume element.
Volume and integrated ratios have the same value
if the medium is homogeneous.

In general the line ratios also depend on $T$
and $x_\mathrm{H}$, but the following analysis is focused
on ratios between lines emitted from the same species (independent
of $x_\mathrm{H}$)
and nearly independent of $T$ (e.~g. the red
$[\rm{SII}]$ ratio).
Assuming a medium with a uniform
electron density $n_{\mathrm{e},0}$, eq. \ref{eq:g12} leads to
$ G_{12} = i_1/i_2 = g_{12}(n_{\mathrm{e},0})$.
Inverting this equation it is possible to determine the density as
$ n_{e,0} = g_{12}^{-1} (G_{12})$.

If the medium is inhomogeneous, the previous formula
may also be applied to give a ``spectroscopic 
electron mean density'' defined as (ABS89):
\begin{equation}
 <n_\mathrm{e}> = g_{12}^{-1} \left[ \frac{\int i_1 dV}
                          {\int i_2 dV} \right].
 \label{eq:spec}
\end {equation}
The volume emissivities for the line $j$ may
be written as (e.~g. ABS89, 
Osterbrock \cite{o89}, Safier \cite{s92}):
\begin{equation}
  i_j = k_j \frac{{n_\mathrm{e}}^2}{1+{n_\mathrm{e}}/n_j}\,,
  \label{eq:ijfit}
\end {equation}
This expression is exact for a two-level atom.
In this case, $n_j=A_{2,1}/(C_{2,1}+C_{1,2})$ is the ``critical density'' 
and $k_j=\frac{1}{4 \pi} h \nu_{2,1} (C_{1,2}/x_{\mathrm{H}}) (n_{\mathrm{s}}/n_{\mathrm{H}}) $,
where $A_{2,1}$ is the Einstein-A coefficient for the transition, 
$C_{1,2}$ and $C_{2,1}$ are the collisional excitation and de-excitation
rates respectively, $h \nu_{2,1}$ is the transition energy, and
$n_{\mathrm{s}}/n_{\mathrm{H}}$
is the population fraction of the species s.
A fit with eq. \ref{eq:ijfit} is also a very good approximation
for many forbidden emission lines, in particular for the red [SII] lines.

Using eq. \ref{eq:ijfit} for replacing $i_j$ in eq. \ref{eq:g12},
we then have:
\begin{equation}
  <n_\mathrm{e}> = \frac{k_1/k_2 -G_{12}}{G_{12}/n_1- k_1/(k_2 n_2)}.
  \label{eq:spdef}
\end {equation}
This equation is next applied to the case of a stellar jet
(see Fig. \ref{fig1}), first considering 
spectroscopic observations with a long slit along the jet axis,
where no spatial information across the jet is available.
The observed line emissivity will be given by the integration 
of the volume emissivity within the jet volume, where
$dV = 2 \pi r dr \Delta z$, being $\Delta z$ the size of 
the beam in the $z$ direction.

If the electron density is represented by a Gaussian radial profile,
then $n_\mathrm{e}=n_0 e^{-r^2/\sigma^2}$, and 
the spectroscopic mean density defined by eq. \ref{eq:spdef} becomes:
\begin{equation}
 <n_\mathrm{e}> = \frac{(n_1-n_2)(n_0-n_\mathrm{ex})-n_1^2 L_1+n_2^2 L_2}
            {n_1 L_1-n_2 L_2},
  \label{eq:nspec}
\end {equation}
where $L_{1,2}=\ln (1+n_0/n_{1,2})/(1+n_\mathrm{ex}/n_{1,2})$,
with $n_0$ the electron density on the jet axis and $n_\mathrm{ex}$
the density evaluated at the jet radius.
This is the same expression derived by ABS89 
for the case of an exponentially decreasing density in an
isothermal stellar atmosphere.

On the other hand, the mean electron density across the jet is given by:
\begin{equation}
  \overline{n}_\mathrm{e} = \frac{\int n_\mathrm{e} dV} {\int dV} = 
  \frac{n_0-n_\mathrm{ex}}{\log \left(n_0/n_\mathrm{ex}\right)},
  \label{eq:nav}
\end {equation}
where the right hand side term is obtained as before assuming a Gaussian 
density profile and $dV = 2 \pi r dr \Delta z$.

The ratio between mean spectroscopic density and 
mean density (i.~e. the ratio between the 
``observed'' and the ``real'' density)
is shown in Fig. 2 as a function of the central electron
density for the [SII] ratio. 
Similar curves were calculated by ABS89 for line ratios 
relevant to stellar atmospheres.

\begin{figure}
\centering
\includegraphics[width=8cm]{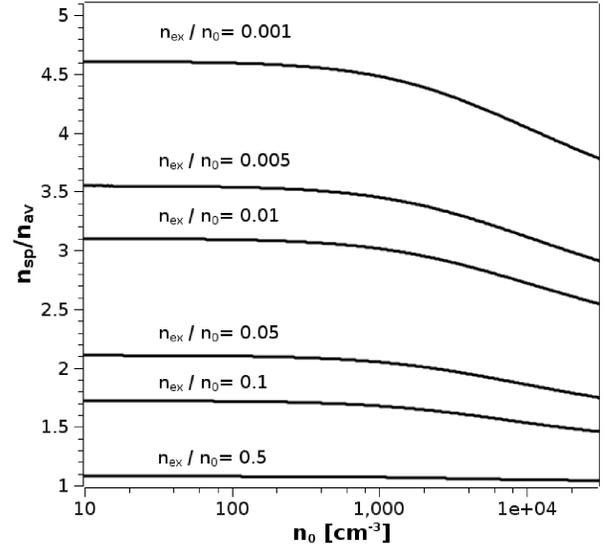}
\caption{Ratio between the spectroscopic $n_{sp}$
         and average density $n_{av}$ as a
         function of the density $n_0$ on the jet axis,
         for different jet cross sections, characterized by
         their limb-to-centre electron density ratio
         $n_\mathrm{ex}/n_0$.}
\label{fig2}
\end{figure}

As can be seen from Fig. \ref{fig2}, the ratio
between the spectroscopic and the mean density increases 
for steeper stratifications of $n_\mathrm{e}$
across the beam of the jet.
At a fixed limb-to-centre electron density ratio
$n_{\mathrm{ex}}/n_0$, the $<n_\mathrm{e}>/\overline{n}_\mathrm{e}$ ratio 
is slowly decreasing.
The spectroscopic mean density is always
larger than the mean density because 
denser regions contribute more to the emission. 
In the low density regime ($n_0, n_{ex} \ll n_j$)
in particular, it is straightforward to show that eq. \ref{eq:nspec}
reduces to $<n_\mathrm{e}>/\overline{n}_\mathrm{e} \approx (2/3) 
(\alpha^2+\alpha+1)/(\alpha^2-1) \ln \alpha$
where $\alpha = n_0/n_{ex}$, and 
$<n_\mathrm{e}>/\overline{n}_\mathrm{e} \approx 2/3 \ln \alpha$ when $n_0 \gg n_{ex}$
as in the upper curves of Fig. \ref{fig2}.

The results shown in Fig. \ref{fig2} are particularly relevant for
stellar jets because an important parameter that is determined  
observationally is the mass flux (and the momentum flux, that has 
the same behaviour, as described below), defined by
$d\dot{M} \sim \rho v r dr $, 
where $\rho$ is the mass density 
($\propto n_\mathrm{H}=n_\mathrm{e}/x_\mathrm{H}$) and v$(r)$ is the jet velocity.
If velocity and ionization fraction have a ``top-hat'' profile in the 
$r$ direction, it may easily be shown that:
\begin{equation}
  \frac{\overline{\dot{M}}}{<\dot{M}>} = \frac{\overline{n}_\mathrm{e}}{<n_\mathrm{e}>},
\end {equation}
where $\overline{\dot{M}}$ is the mass flux and the 
$<\dot{M}>$ is the one determined using the spectroscopic
mean density.
Therefore, values of mass and momentum-flux represent 
just an upper limit to the real values (see e.~g. Cabrit \cite{c02}).

\subsection{Convolution effects for a resolved jet cross section}

We now consider the case where information on the emission
profile across the jet is available.
The volume and integrated emissivities $i(r)$
and $I(x)$ are related to each other by the Abel transform:
\begin{equation}
  I(x) = 2\int_x^{R} \frac{i(r)r dr}{\sqrt{r^2-x^2}},
  \label{eq:abel}
\end {equation}
where $R$ is the jet radius and $x$ is the projected separation
between a point in the observed image and the projection of the
jet axis on the plane of the sky.

In the case of an homogeneous medium $i=i_0$ and eq. \ref{eq:abel} leads to
$I(x) = 2 i_0 \sqrt{R^2-x^2}$.
To show the effect of an inhomogeneous medium on plasma diagnostics,
in the following we present results obtained numerically integrating
eq. \ref{eq:abel} for different $n_\mathrm{e}$, $T$, $x_\mathrm{H}$ profiles.

\begin{figure}
  \centering
    \includegraphics[width=9cm]{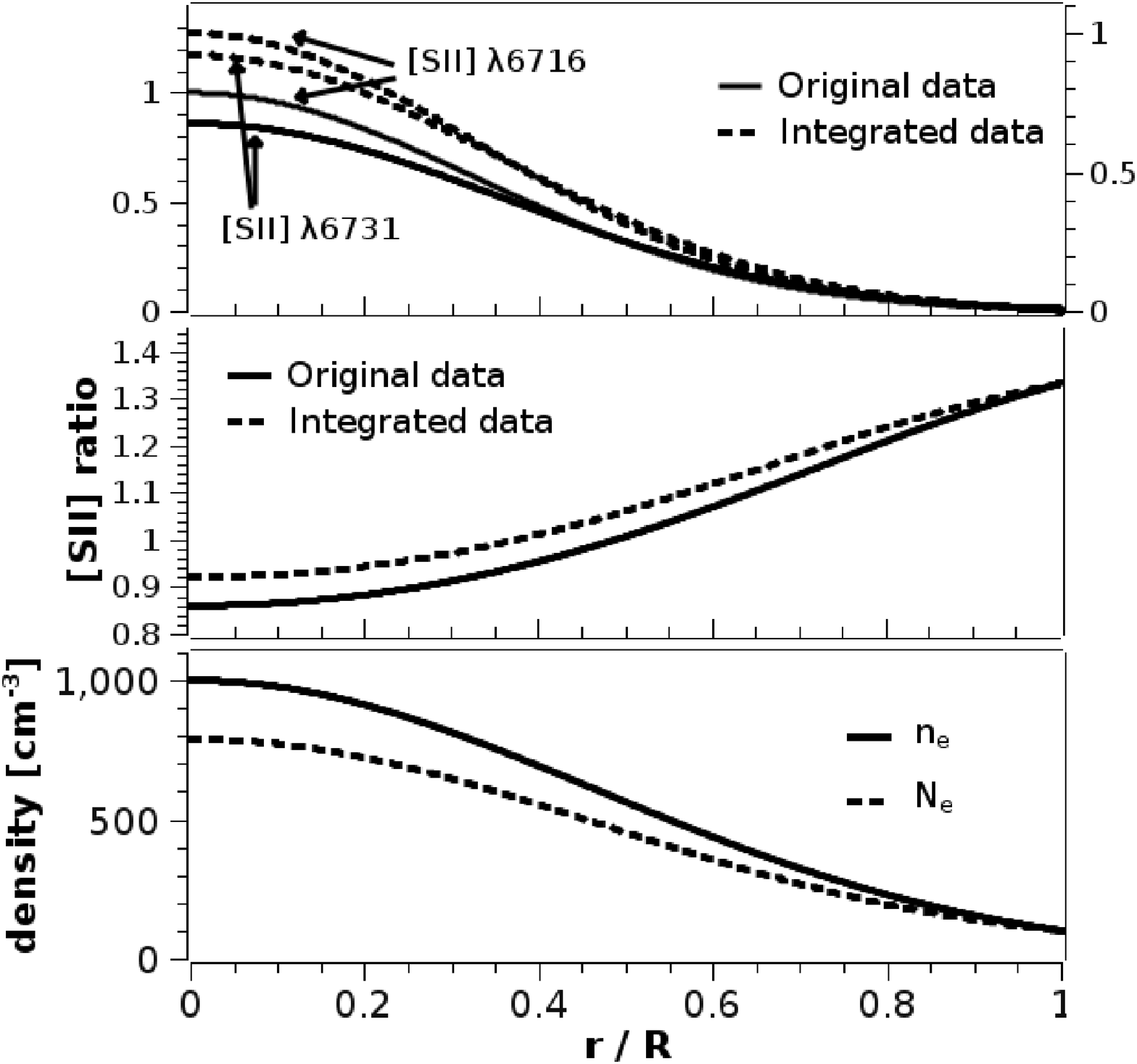}
\caption{Effect of the convolution on inhomogeneous jets.
   The dotted lines represent integrated quantities, while
   the continuous lines represent volume quantities.
   The electron density has a Gaussian profile ranging between
   10$^3$ cm$^{-3}$
   on the axis and 100 cm$^{-3}$ at the jet radius.
   The temperature and the ionisation fraction are
   assumed to have constant values of 10$^4$ K and 0.1, respectively.
   \emph{Top}: $[\rm{SII}] \lambda6716, \lambda6731$ integrated
      and volume intensities. 
   \emph{Centre}: $[\rm{SII}] \lambda6716/ \lambda6731$ integrated
   and volume ratio.
   \emph{Bottom}: volume and integrated electron density.
     }
  \label{fig3}
\end{figure}

Assuming a profile for $n_\mathrm{e}$, $T$, $x_\mathrm{H}$,
for each radial position the synthetic emissivities in the 
$[\rm{SII}] \lambda\lambda6716, 6731$, $[\rm{NII}] \lambda6583$, 
and $[\rm{OI}] \lambda6300$ lines are calculated,
solving the statistical equilibrium equations 
using a 5-level atom (with atomic parameters from Mendoza \cite{m83}).
These spatially dependent emission coefficients are then used
to numerically integrate eq. \ref{eq:abel}.
Finally, the physical parameters are determined
from the integrated emission line intensities
using the BE method (see Appendix \ref{apB}).

The results obtained assuming a Gaussian profile
for $n_\mathrm{e}$, and constant $T$ and $x_\mathrm{H}$
cross sections are shown in Fig. \ref{fig3}.
The volume and integrated emissivities of $[\rm{SII}] \lambda6716$
and $[\rm{SII}] \lambda6731$ versus $r/R$ are shown in
the top panel, both normalised to their values at $r=0$.
The [SII] ratio is also shown in the central panel of Fig. \ref{fig3}
while the volume and integrated electron densities are shown in the 
lower panel.

The [SII] ratio determined using the integrated emissivities is higher than
the original ratio (continuous and dotted line of Fig. 3, respectively).
The difference between $n_\mathrm{e}$ and $N_\mathrm{e}$ 
increases from 0 at the external radii to 
 approximately 20\% toward the jet axis.

\begin{figure}
\centering
\includegraphics[width=8cm]{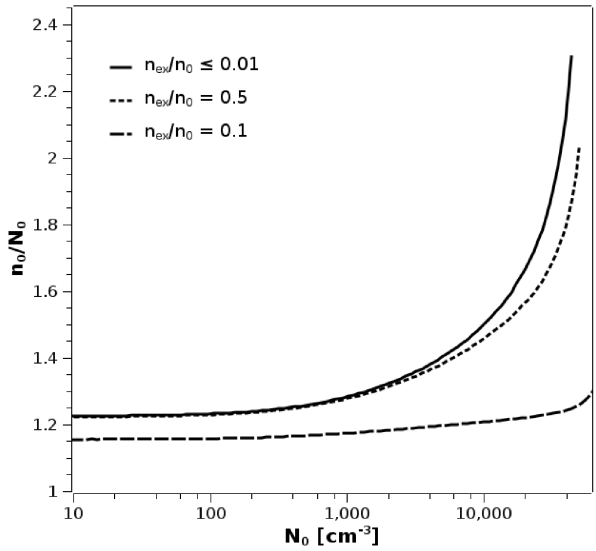}
\caption{Ratio between convolved
   and volumetric densities for different
   values of $n_0/n_\mathrm{ex}$.}
\label{fig4}
\end{figure}

Fig. \ref{fig4} shows the ratio $n_0/N_0$ 
between the original and convolved
values of the electron density on the jet axis
versus the convolved electron density $N_0$,
for different values of $n_\mathrm{ex}/n_0$.
As is evident from the figure, the convolution effect increases 
when the observed density approaches the critical
density (of $\sim 10^4$ cm$^{-3}$) for the $[\rm{SII}]$ ratio.
For high density stratifications, the 
$n_0$/$N_0$ ratio converges to a maximum ratio, which depends
on the value of $N_0$.

\begin{figure}
\centering
\includegraphics[width=8cm]{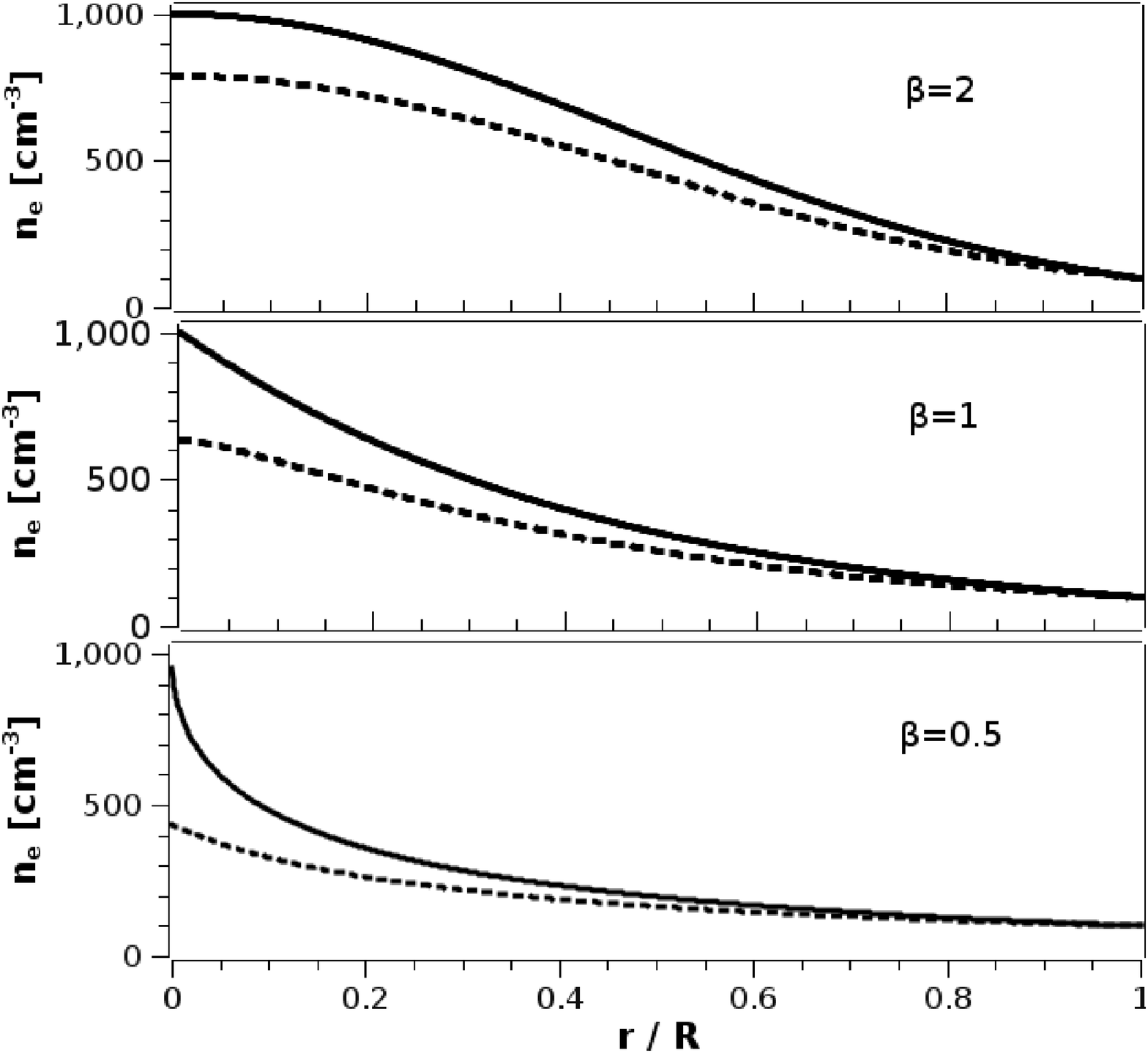}
    \caption{The dotted lines represent convolved quantities, and
           the continuous lines represent volume quantities.
           The curves correspond to the integrated and volume densities,
           obtained using the following initial profiles:
           $n \sim 10^{-(r/R)^\beta}$,
           $T=10000$~K, $x_\mathrm{H}=0.1$ 
           where $\beta=2,1,0.5$ in the upper, central and lower
           panel respectively.
   The volume electron density is 1000 cm$^{-3}$
   on the axis and 100 cm$^{-3}$ at the jet radius.
         }
\label{fig5}
\end{figure}

Fig. \ref{fig5} shows the results obtained for different
profiles for $n_\mathrm{e}$ and top hat cross sections for
$T$ and $x_\mathrm{H}$.
The differences between $n_0$ and $N_0$ goes up
to $\sim$ 60\% for the most stratified case (bottom
panel).

\begin{figure}
  \centering
    \includegraphics[width=9cm]{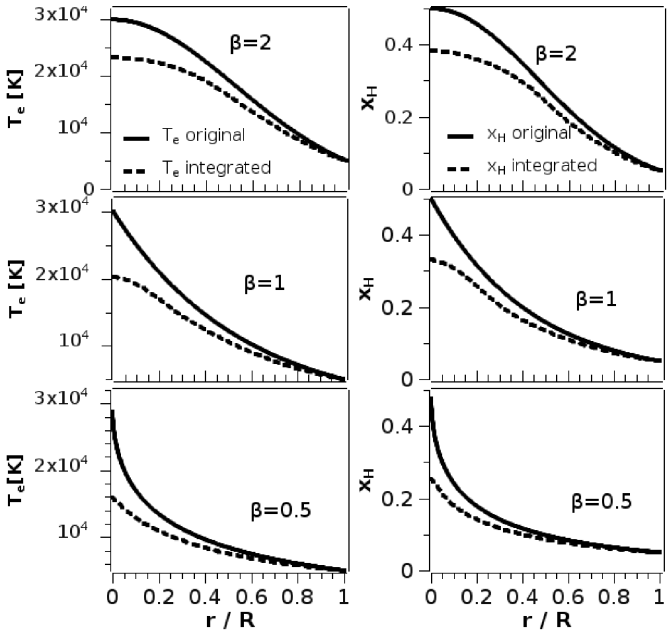}
    \caption{Effect of the convolution on inhomogeneous jets.
   The dotted lines represent integrated quantities, and
   the continuous lines represent volume quantities.
   Thin lines correspond to the ionisation fraction, and
   thick lines to the temperature.
   The curves are obtained using the following profiles:
   $T_\mathrm{e}, x, n_ e \sim 10^{-(r/R)^\beta}$,
           where $\beta=2,1,0.5$ in the upper, central and lower
           panel, respectively.
         }
  \label{fig6}
\end{figure}

Fig. \ref{fig6} shows the results
obtained for different initial stratifications 
in $n_\mathrm{e}$, $T$ and $x_\mathrm{H}$,
corresponding to radial profiles with the form 
$ \sim 10^{-(r/R)^{\beta}}$
where $\beta=2,1,0.5$
for the top, centre and bottom panels
of Fig. \ref{fig5} and \ref{fig6} respectively.

It is clear from Fig. \ref{fig6} that effects similar 
to the one observed for $n_\mathrm{e}$ in Fig. \ref{fig3}
are present also for $T$ and $x_\mathrm{H}$. 
Additionally, Fig. \ref{fig5} and \ref{fig6} show 
that convolution effects increase with the
plasma stratification (as expected).


\section{Data reconstruction}

\subsection{Reconstruction of the jet structure using a Gaussian fit to the observed intensities}

The volume emission coefficient projected on the sky plane
is also weighted on the beam size $\Delta x \Delta z$.
Therefore, the observed emissivity $I(x_i)$ is given by:
\begin{equation}
  I(x_i) = 2 \Delta z \int_{x_{i-1/2}}^{x_{i+1/2}} dx
     \int_x^{R} \frac{i(r)r dr}{\sqrt{r^2-x^2}},
  \label{eq:abelintp}
\end {equation}
where $i=-N, ...,0, ..., N$, there are $2N+1$ observational values 
(with $N=0,1,2,$ \dots),  and $i_{\pm (N+1/2)}=R$.
All the relevant quantities are assumed to have negligible variations
in the $z$ direction.

A simple scaling between $i(r)$ and $I(x)$ 
may be obtained in the case of an inhomogeneous medium,
using the linearity of the Abel transform of a
Gaussian distribution.
In fact, if the volume emissivity has the form
$i(r) = i_0 e^{-r^2/\sigma^2}$, eq. \ref{eq:abel} leads to:
\begin{equation}
  I(x) = \sqrt{\pi} \sigma i(x) \mathrm{erf}
       \sqrt{  \ln{\frac{i_0}{i_{\mathrm{ex}}}}
       \left[1-\left(\frac{x}{R}\right)^2 \right] },
 \label{eq:abelgaussian}
\end {equation}
where 
\begin{equation}
  \mathrm{erf}(x) = \frac{2}{\sqrt{\pi}} \int_0^x e^{-t^2} dt
 \label{eq:error}
\end {equation}
is the error function, and $i_0$ and $i_{\mathrm{ex}}$ are the
volume emissivity at the jet centre and limb, respectively.

The property of the error function: $\rm{erf}(x) \approx 1$
when $x \gtrsim 1$ implies that
\begin{equation}
  I(x) \approx \sqrt{\pi} \sigma i(x)
 \label{eq:abelsimpl}
\end {equation}
if  $x \lesssim R$ and $i_0 \gg i_{\mathrm{ex}}$.

Using $I(x)$ as given by eq. \ref{eq:abelsimpl}, eq. \ref{eq:abelintp}
becomes:
\begin{equation}
  I(x_i) = \pi \sigma^2 i_0 \left(\rm{erf}\frac{x_{i+1/2}}{\sigma}
                                          -\rm{erf}\frac{x_{i-1/2}}{\sigma}
   \right)\Delta z ,
  \label{eq:noaprox}
\end {equation}
If only one value of $I$ is observed 
($x_{i+1/2}=-x_{i-1/2}=R$),
the integrated emissivity will be simply given by:
\begin{equation}
  I = \pi \sigma^2 i_0 \left( 1-\mathrm{e}^{-R^2/\sigma^2}\right)\Delta z 
   \approx \pi \sigma^2 i_0 \Delta z .
  \label{eq:aproxi}
\end {equation}
%
%
Given eq. \ref{eq:aproxi},
the relation between the integrated ratio $G$
and the corresponding volume ratio $g$, in the 
case of Gaussian profile emissivities, is given by:
\begin{equation}
   G = \eta g  \qquad   \rm{where} \qquad
   \eta = \left( \frac{\sigma_1}
                      {\sigma_2} \right) ^2.
 \label{eq:g}
\end {equation}
If $\sigma_1 \neq \sigma_2 $, the integrated and volume ratios
will be different. Therefore, the physical parameters determined
from the \emph{integrated ratio} will be different
with respect to the one obtained
using the \emph{volume ratio}.
Furthermore, the simple scaling given by eq. \ref{eq:g} may be used
to reconstruct the volume intensities given the integrated intensities and 
the values of $\sigma_1$ and $\sigma_2$, and from there to 
determine the radial dependence of the physical parameters. 

We first show how to apply eq. \ref{eq:g} to the [SII] ratio
to determine $n_\mathrm{e}$, 
and later the results will be generalised to more line ratios, to 
determine all of the physical parameters.

\begin{figure}
\centering
\includegraphics[width=8cm]{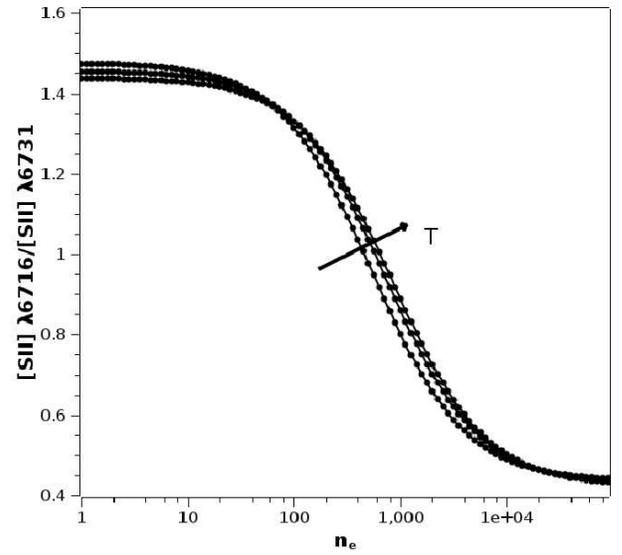}
\caption{ $\mathrm{[SII]} \lambda6716/ \lambda6731$ ratio (points)
          and fits (lines) with eq. \ref{eq:fit}, corresponding
          to different temperatures. The fitting parameters
          are the following:
             $n_0=$ 1841.89, 2457.4, 2883.9 cm$^{-3}$,
             $a_1=$ 0.678, 0.687, 0.695,
             $a_2=$ 2.295, 2.333, 2.336
           for $T=5000, 10000, 15000$ K, respectively.
                 }
\label{fig7}
\end{figure}

A very good approximation to the relation between the [SII] ratio and 
$n_\mathrm{e}$ may be obtained from eq. \ref{eq:spdef} and is given by:
\begin{equation}
   g = \frac{n_\mathrm{e}+n_0}{n_\mathrm{e} a_2 + n_0 a_1},
  \label{eq:fit}
\end {equation}
where the parameters $n_0$, $a_1$ and $a_2$ are in general
temperature dependent.
This formula is exact for a two-level atom, and 
in general is just a fit to the $g=g(n_\mathrm{e})$
curve at constant $T$ and $x_\mathrm{H}$.
Fig. \ref{fig7} shows the [SII] ratio and the fit calculated
using eq. \ref{eq:fit}. The fitting parameters are reported in the 
Figure caption as a function of temperature.

\begin{figure}
\centering
\includegraphics[width=8cm]{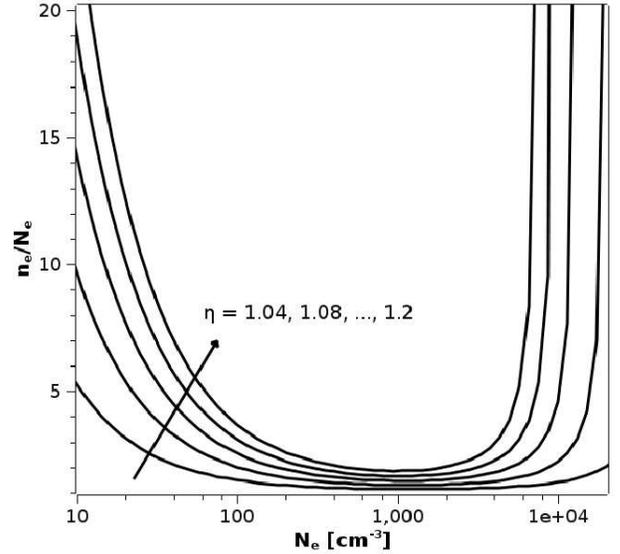}

\caption{Ratio of volume and integrated electron densities
         as a function of the integrated electron density for different
         values of $\eta$.}
\label{fig8}
\end{figure}

In eq. \ref{eq:fit}, $n_\mathrm{e}$ and $g$ represent the volume 
electron density and ratio respectively.
A similar relation will also link the integrated
quantities $N_\mathrm{e}$ and $G$:
\begin{equation}
   G = \frac{N_\mathrm{e}+n_0}{N_\mathrm{e} a_2 + n_0 a_1}.
  \label{eq:FIT}
\end {equation}
%
%
Using eq. \ref{eq:g}, \ref{eq:fit} and \ref{eq:FIT}
it is possible to obtain a relation between $n_\mathrm{e}$
and $N_\mathrm{e}$:
\begin{equation}
  n_\mathrm{e}=n_0 \frac{N_\mathrm{e} (\eta a_2 - a_1)+n_0 a_1(\eta-1)}
                        {N_\mathrm{e} a_2 (1-\eta)+n_0 (a_2 -\eta a_1)}.
  \label{eq:neNe}
\end {equation} 
Fig. \ref{fig8} shows the ratio between the 
volumetric and the observed densities $n_\mathrm{e}/N_\mathrm{e}$
(obtained from eq. \ref{eq:neNe}) as function of $N_\mathrm{e}$, 
for different values of $\eta$.
For electron densities $N_\mathrm{e} \gtrsim 5000$ cm$^{-3}$ and
$N_\mathrm{e} \lesssim 100$ cm$^{-3}$, the $n_\mathrm{e}/N_\mathrm{e}$
ratio becomes increasingly larger.
These density ranges correspond to the high and low-density regimes
(see Fig. \ref{fig7}).
Also, for densities around $10^3$ cm$^{-3}$, the differences may be as 
high as a factor of 2 or 3, depending on the value of $\eta$.

This procedure may naturally be generalised to include the 
calculation of $T$ and $x_\mathrm{H}$:
\begin{itemize}
\item for each line with spatial information 
across the jet, the value of $\sigma$ is obtained with a Gaussian fit 
to the observed emission profile.
\item The volumetric line ratios are reconstructed using eq. \ref{eq:g} as
$g_{1,2}=i_1/i_2=\sigma_2/\sigma_1 G_{1,2}$.
\item The reconstructed ratios are used to determine the physical
parameters using any of the available methods (e.~g. the BE method, 
see Appendix \ref{apB}).
\end{itemize}

If the jet is observed through a narrow slit placed along the
main jet axis, eq.
\ref{eq:noaprox} has to be used instead of eq. \ref{eq:aproxi}.
Therefore, $\eta$ will be defined in this case as:
\begin{equation}
   \eta = \left( \frac{\sigma_1}
                      {\sigma_2} \right) ^2
         \frac{\mathrm{erf} \left(\Delta x/\sigma_1 \right)}
              {\mathrm{erf} \left(\Delta x/\sigma_2 \right)},
 \label{eq:eta}
\end {equation}
where $\Delta x$ is the beam size along the $x$-direction.

\subsection{Reconstruction of the jet structure using tomographic techniques}

The Abel transform (eq. \ref{eq:abel}) has been largely studied
and applied in science, and complex methods were developed
to solve it (e.~g. Craig \& Brown \cite{cb86}).
The analytical solution of the Abel transform is also well known:
\begin{equation}
  i(r) = -\frac{1}{\pi} \int_r^{R} \frac{d I}{dx}
             \frac{dx}{\sqrt{x^2-r^2}} . 
 \label{eq:ilp}
\end {equation}
The determination of $i(r)$ represents an ``ill-posed''
problem in the sense that small noise in $I(x)$ produces
large errors in the determination of $i(r)$, due to the
numerical derivative present in the integral.
Any numerical method solving eq. \ref{eq:ilp} using noisy observed data
is therefore necessarily unstable.

A method that is less affected by numerical instabilities is obtained 
by fitting the data by a smooth function and using eq. \ref{eq:ilp}
to invert the fitted curve (e.~g. Simonneau \cite{s93}).
In this context, we use the multi-Gaussian method developed by
Bendinelli (\cite{be91}), which has been mainly used for deprojection
of galaxy surface brightness distribution
(e.~g. Bendinelli \cite{be91}, Emsellem et al. \cite{e94},
Bendinelli \& Parmeggiani \cite{be95}, Cappellari \cite{ca02}).
This method basically consists of producing a multi-Gaussian expansion fit
of the data, and inverting the Gaussian series.
Actually, this approach becomes very simple once the fit has been obtained, 
due to the already mentioned linearity of the Abel transform when applied
to a Gaussian distribution.
This implies that for a fit to a data series with a sum
of Gaussians (centred on the same point):
\begin{equation}
  I(x) = \sum_{i=1}^n a_i e^{-x^2/\sigma_i^2} , 
  \label{eq:irec1}
\end {equation}
the Abel transform may be inverted analytically (eq. \ref{eq:ilp}) to give:
\begin{equation}
  i(r) = \sum_{i=1}^n \frac{1}{\sqrt{\pi}}
\frac{a_i}{\sigma_i} e^{-r^2/\sigma_i^2} ,
  \label{eq:irec}
\end {equation}
where $r$ is the distance from the jet axis.

\subsection{Application to HH30}

We now apply the techniques derived above to the HH30 jet.
HH30 is an ideal candidate for this analysis.
It moves nearly on the sky plane,
has a clear side-to-side symmetry in the region close to the
central star, and the cooling region
is resolved spatially with Hubble Space Telescope (HST) observations
(e.~g. Burrows et al. \cite{b96}, Ray et al. \cite{r96}, Bacciotti et al. \cite{ba99},
HM07).
The data used for this analysis were taken by HST using the slit-less 
spectroscopy technique, and presented by HM07.
Using this powerful technique, the observations by HM07 
resolve the cross section of the HH30 jet with $\sim 10-20$ pixels.

One position along the jet axis is considered here, at an angular
separation of $3.9\arcsec$ from the HH30 source,
corresponding to a region close to the central star-disk system.

\begin{figure}
  \begin{center}
    \includegraphics[width=7cm]{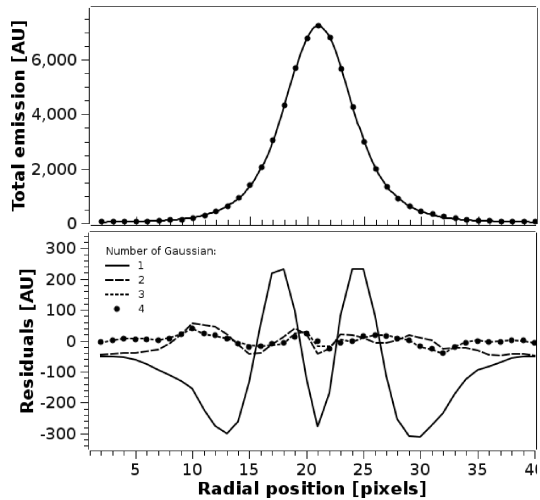}
    \caption{\emph{Upper panel}: Observed line emissivities (points)
               and multi-Gaussian fit as a function of the 
               position across the jet using 3 Gaussian curves.
             \emph{Lower panel}: Residuals for a different 
               number of multi-Gaussian fits.}
    \label{fig9}
  \end{center}
\end{figure}

First, to calculate the centre $x_0$ of the Gaussian fit,
i.~e. the projection of the jet axis on the plane of the sky,
all the observed profiles are added.
The obtained emissivity is then fitted with an increasing number of Gaussians.
The values of $x_0$ and the variance $\sigma$ obtained using up to 
5 Gaussians are shown in Tab. \ref{tab1}.
Fig. \ref{fig9} shows the data and a fit with 3 Gaussians (top),
and the residual (bottom).
From Tab. \ref{tab1} and Fig. \ref{fig9} it is evident that a very
good approximation to the data is obtained already with 3 Gaussians.

  \begin{table}
      \caption{Data fit with Gaussian curves}
\label{tab1}      
\centering                          
\begin{tabular}{c c c}        
\hline\hline                 
Gaussian Number & $x_0$ & $\sigma$\\    
\hline                        
            1     & 20.97757  &    153.4     \\
            2     & 20.98055  &     30.62    \\
            3     & 20.98054  &     16.42    \\
            4     & 20.98061  &     15.71    \\
            5     & 20.98061  &     15.69    \\
\hline                                   
\end{tabular}
\end{table}

\begin{figure}
  \begin{center}
    \includegraphics[width=7cm]{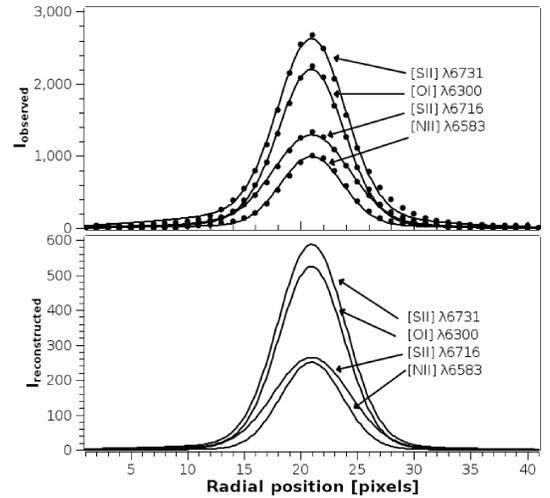}
    \caption{\emph{Upper panel}: Observed line emissivities (points)
               against multi-Gaussian fit as a function of the 
               position across the jet.
             \emph{Lower panel}: Reconstructed line emissivity.}
    \label{fig10}
  \end{center}
\end{figure}

\begin{figure}
  \begin{center}
    \includegraphics[width=8cm]{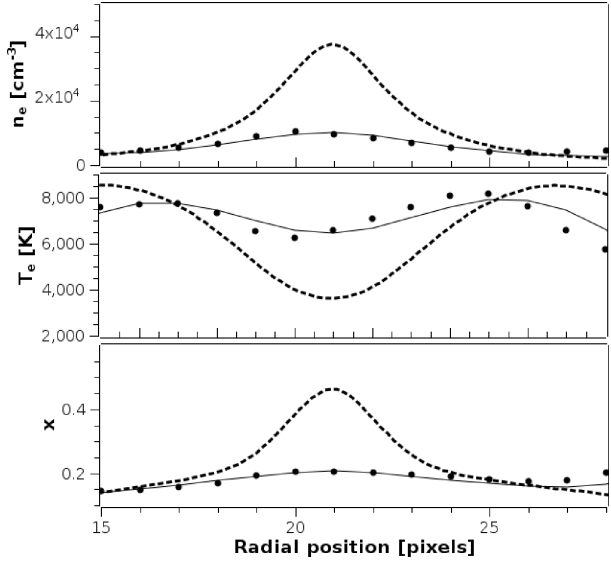}
    \caption{Values inferred from observations (points), fitted values 
             (with a multi-Gaussian, continuous line) and
             reconstructed values (dotted line) of electron density, temperature
             and ionisation fraction for the upper, central and lower panels
             respectively.}
    \label{fig11}
  \end{center}
\end{figure}

The results obtained are shown in Fig. \ref{fig10}
and Fig. \ref{fig11}.
Each emissivity profile is fitted with a multi-Gaussian
fit, and using eq. \ref{eq:irec} the intensities
as a function of the radial position, and the corresponding
ratios are reconstructed.

Fig. \ref{fig10} shows the observed data, the multi-Gaussian fit (top)
and the reconstructed data (bottom).
The ratios calculated using the reconstructed data differ from
the original, observed ratios (see for example the 
[SII] $\lambda6716$ curves with respect to the [NII] in the upper
and lower panels of Fig. \ref{fig10}).

The physical quantities are shown in Fig. \ref{fig11}.
In the upper panel $n_\mathrm{e}$ is shown.
The points correspond to the observed values, while the 
continuous and dotted lines correspond to the multi-Gaussian fit
and the reconstructed electron density values, respectively.
The central and lower panels show the temperature and the ionisation
fraction.

Some interesting features may be noted from  Fig. \ref{fig11}.
First at all, it is evident that HH30 does not have a top hat
cross section. The reconstructed $n_\mathrm{e}$, $T$ and $x_\mathrm{H}$
cross sections are much steeper than the measured cross sections.
The reconstructed electron density and ionisation fraction cross sections
have strong, on-axis peaks, and the temperature has an on-axis valley.
The total densities ($n_\mathrm{H}=n_\mathrm{e}/x_\mathrm{H}$) 
are larger by a factor of two and four on the jet axis with respect to the 
jet radius, for the inferred from observation and reconstructed case respectively.
The fact that we have lower temperatures in the denser regions is
qualitatively consistent with the stronger cooling that one would expect
in higher density regions.
The fact that the ionisation fraction is
higher in the denser regions, however, is an effect that escapes simple
qualitative arguments, and should be compared directly with predictions
of jet formation and propagation.

It is important to note that the original data
were not deconvolved with the PSF (with a width of $\approx 2.5$ pixels),
and we may expect a much stepper profiles using deconvolved data.
Also, the jet presents some degree of asymmetry (see e.~g.
the wings of the observed temperature in the central panel 
of Fig. \ref{fig11}).
All of these effects, together with a complete study of the two-dimensional
structure of the HH30 jet are left for a second paper.

It should be noted that the results shown in Fig. 11 are affected
by the fact that the [SII] ratio is close to the high density
regime (see Fig. 8).
Other line ratios (see the discussion by HM07) should be used to 
deconvolve the jet beam closed to the source.


\section{Summary and discussion}

In this paper we have studied the influence of inhomogeneities on
the electron density, temperature
and ionisation fraction values derived from observed line ratios.
Additionally, we have presented possible methods to analyse the data
obtained from high angular resolution imaging of stellar jets.

Simple scalings are possible when a minimum amount of information
is available. This minimum information is one value of emission
line intensities
for each position along the jet, together with a determination of
the jet width $\sigma$ for \emph{all} of the different lines used for
the plasma diagnostics.

Bacciotti et al. (\cite{b00}) observed DG Tau with
7 narrow slits parallel to the main jet axis, 
and at different positions across the jet,
while Coffey et al. (\cite{c07}) used 
one slit placed perpendicular to the main jet axis.
In both cases, they obtained spatial and kinematic information
across the jet axis.
The approach presented in \S 3.1 can be easily applied
to these data, calculating the average
physical parameters \emph{and}
jet velocity as a function of position along the jet.
The determination of the jet velocity
has not been explicitly discussed in the present paper,
but similar methods to the ones that we have discussed above can be
used in this context.

We have also shown that a more complex 
tomographic reconstruction (see \S 3.2) can be used to reconstruct
the cross section of the flow, obtaining a complete
description of the three dimensional structure of the jet.


The main limitations for the application of these techniques
is that the jet axis has to lie close to the plane of the sky.
However, it would be possible (but more complex)
to reconstruct the 3D structure of the flow for arbitrary
orientations of the jet axis.
Additionally, the jet has to be nearly axisymmetric.
Also, the applicability of the present deconvulation method is limited
to the nearly axysymmetric region of the jet close to the source.
Farther away from the source, in fact, the interaction of the jet with the 
ambient medium produces asymmetries in the intensity cross-sections.

Finally, while this technique is currently applicable to a few observed 
objects, the use of future high resolution instruments of the new generation
of telescopes will make it possible to apply this technique to a larger sample
of objects.

In this paper, preliminary application of our
method to reconstructing the cross section of the HH 30 jet
has been presented. In a second
paper, we will describe a much more complete application of our method
using observations of HH 30 and of other HH jets.


\appendix

\section{BE method}
\label{apB}

The emission coefficient (units: erg s$^{-1}$ cm$^{-3}$ sr$^{-1}$) obtained by the transition
from the $i$ to $j$ levels is given by:
\begin{equation}
  i_{i,j}= \frac{1}{4 \pi} h \nu_{i,j} A_{i,j} \frac{n_i}{n_{A_i}} 
          \frac{n_{A_i}}{n_A} \frac{n_A}{n_\mathrm{H}} n_\mathrm{H},
  \label{eq:int}
\end {equation}
where $h \nu_{i,j}$ is the transition energy, $A_{i,j}$ is the 
spontaneous Einstein coefficient, $n_i/n_{A_i}$, 
$n_{A_i}/n_A$ and $n_A/n_\mathrm{H}$ are the 
excitation fraction, the ionisation fraction
and the population fraction of the considered species.

The BE method (from BE99) uses a series of line ratios to determine 
 $n_\mathrm{e}$, $T$, $x_\mathrm{H}$:
 ${\rm{[SII]} \lambda6731}$/${\rm{[SII]} \lambda6716}$,
 ${\rm{[OI]} \lambda6300}$/${\rm{[SII]} \lambda6716+6731}$,
 ${\rm{[NII]} \lambda6583}$/${\rm{[SII]} \lambda6716+6731}$.
The [SII] ratio depends on $n_\mathrm{e}$ and $T$, while the [OI]/[SII]
and [NII]/[SII] ratios depend also on n$_{\rm{OI}}$/n$_{\rm{SII}}$ and 
n$_{\rm{NII}}$/n$_{\rm{SII}}$ respectively.
Additionally, nitrogen and oxygen ionisation fraction are determined
assuming charge exchange equilibrium with hydrogen.
Therefore n$_{\rm{OI}}$/n$_{\rm{SII}}$ and n$_{\rm{NII}}$/n$_{\rm{SII}}$
become a function of the ionisation fraction and 
n$_{\rm{O}}$/n$_{\rm{S}}$ and n$_{\rm{N}}$/n$_{\rm{S}}$
respectively.
Finally, the sulphur is supposed to be all single ionised (n$_{\rm{SII}}$ = n$_{\rm{S}}$)
because the photoionisation rate due to diffuse UV radiation 
(e.g. Tielens \cite{tie05}, pp. 267)
is much larger than the SII radiative recombination rate.

To calculate the physical quantities, we choose arbitrary 
values of temperature and ionisation fraction, and 
invert the [SII] ratio deducing the electron density. Furthermore we use the N/S ratio
and the electron density derived previously to determine the ionisation
fraction, and finally we use the O/S ratio to find the electron temperature.
Finally, we iterate until convergence using the new values of $T$ and 
$x_\mathrm{H}$.

The atomic parameters used to calculate collisional ionisation, radiative +
dielectronic recombination and charge exchange
coefficients are from Cox (\cite{c70}), Aldrovandi \& P\'equignot (\cite{a73, a76}) and 
Osterbrock (\cite{o89}).


\begin{acknowledgements}
The authors kindly acknowledge Edith Salado Lopez for the elaboration of Fig. 
\ref{fig1}, and Linda Podio for useful discussions. A special thanks to Pat Hartigan
for sharing his HH30 data.
FDC acknowledges Catherine Dougados for an inspiring lecture
on the effects of inhomogeneities
on stellar jets during the V Jetset School.
FDC also acknowledges support of the European
Community's Marie Curie  Actions - Human Resource and Mobility within
the JETSET (Jet Simulations, Experiments and Theory) network under
contract MRTN-CT-2004 005592.
AR acknowledges support from
the DGAPA (UNAM) grant IN108207, from the CONACyT grants 46828-F and
61547, and from the ``Macroproyecto de Tecnolog\'\i as para la Universidad
de la Informaci\'on y la Computaci\'on'' (Secretar\'\i a de Desarrollo
Institucional de la UNAM).
Finally, the authors thank the anonymous referee for useful comments.
\end{acknowledgements}



\end{document}